# Landscape of grain boundary migration in polycrystals

Jianfeng Hu[1]*

[1] School of Materials Science and Engineering, Shanghai University; Shanghai 200444, China.

*Correspondence Email: jianfenghu@shu.edu.cn

**Abstract**

Grain boundary (GB) migration is a pivotal process that significantly impacts the development of microstructures, thereby influencing the practical performance of polycrystalline materials. Recent advances in 3D experimental techniques have revealed conflicts between observed GB migration behaviors and classical theoretical models. These contradictions raise two fundamental questions, namely, whether GB migration is linearly related to curvature, and how GB energy affect GB migration? Here, we provide a comprehensive analysis of GB migration dynamics in polycrystals and resolve these conflicts within a theoretical framework. Unexpectedly, in a polycrystalline system, the range of GB migration velocities shows little correlation with the magnitude of its curvature. The extent of the influence of GB energy on GB migration is revealed to mostly depend on GB step energy. Finally, a more general GB migration formula is derived to incorporate various driving forces beyond curvature.

## 1. Introduction

The formed microstructure resulted from grain growth dominates the performances of polycrystalline materials [1–5]. The control of grain size is especially important to obtain and maintain the unique properties of nanograined materials [5–7], and grain size is also related with the mechanical properties of materials as the description of the well-known Hall-Petch relationship [8]. There are two different pathways with a competitive relationship for the growth of grains: conventional GB migration mediated by atomic diffusion, and ordered coalescence of crystallites [9]. In most cases, grain growth is completed through GB migration that determines the growth behavior of grains due to the stringent activation conditions of ordered coalescence of crystallites [10]. Currently, the reaction rate theory is the most widely accepted model for GB migration, which suggests the simple concept of transferring atoms across a boundary driven by the product of the curvature and GB energy [11]. The formula of GB migration in reaction rate theory as follows,

$$v = M\kappa\gamma \qquad \qquad 1)$$

where the mobility $M$ is an Arrhenius-type coefficient. $\kappa$ and $\gamma$ are the curvature and GB energy, respectively. This model has been experimentally verified in bicrystals [11]. Meanwhile, based on this migration model, several classical grain growth formulas have been proposed, such as the von Neumann relation [12] and Hillert's theory [13], which can phenomenologically describe the self-similar behavior of the mono-modal grain size distribution during normal grain growth (NGG) processes. However, these grain growth formulas cannot adequately describe or interpret the bimodal grain size distribution observed during abnormal grain growth (AGG) [14]. Furthermore, recent 3D experimental results on GB migration in polycrystalline systems have challenged the classical Eq. 1 in reaction rate theory [15,16]. Except for the reaction rate theory, several mechanisms have been proposed to describe GB migration in the past few decades, such as the terrace-ledge-kink model [17] and the disconnection model [18,19]. Unlike Eq. 1, these models incorporate parameters of specific GB structures like the step structure and dislocation, which may contribute to a comprehensive understanding of the detailed migration behavior of individual GB with crystalline structure at the atomic scale. However, due to the complex GB structure parameters involved in these equations, as well as the coexistence of various complex and diverse types of GB structures (like liquid-like GBs and crystalline GBs) in a polycrystalline system, it is difficult to directly validate these GB migration models through statistically experimental observations of GB migrations. Consequently, these models cannot be utilized to derive formulae for describing grain growth behaviors.

More recently, a novel curvature-driven GB migration formula has been proposed [14]. Compared to the classical reaction rate theory, this model introduces two new parameters, namely GB step energy and curvature-distribution-related variable, into the equation as the critical variables governing GB migration. GB step energy is closely associated with the intrinsic GB migration barrier. Given that GB step energy, like GB energy, is an intrinsic property of all types of GBs in materials, the newly formulated GB migration equation, analogous to Eq. 1, can thus be generally applied to characterize the migration behavior of various GBs in polycrystalline materials and serve as a basis for deriving the associated grain growth equations. A general grain growth theory based on this migration model has been shown to successfully explain the general behaviors of grain growth, including normal, abnormal and stagnant growth of grains [20]. Therefore, this GB migration formula shows promise in describing the general behavior of GB migration in polycrystalline materials.

Recently, the significant advancements in high-energy diffraction microscopy (HEDM) and diffraction contrast tomography (DCT) technologies enable the direct and nondestructive observation of 3D GB migration in polycrystalline materials [16]. GB migration experiments have been reported in various polycrystalline systems, such as polycrystalline Ni [15], polycrystalline Fe [21–23], Al-Cu alloy [24] and SrTiO$_3$ ceramics [25]. These have allowed for direct experimental verification of current theoretical concepts and models for GB migration. However, 3D GB migration studies revealed contrasting behaviors in different polycrystalline materials: polycrystalline Fe annealed at 800°C exhibited classical behavior where the average GB migration velocity showed a linear correlation with curvature and independence from GB characteristics [21,22]. In contrast, both polycrystalline Ni at 800°C [15] and Fe at 600°C [23] demonstrated no statistically linear correlation between GB migration velocity and curvature, while showing significant dependence on GB crystallographic parameters. These contradictions raise two fundamental questions: (1) whether GB migration exhibits a linear relationship with curvature as shown in Eq. 1, and (2) how GB energy influences the migration process. Therefore, there requires a new GB migration model rather than the classic models to clarify the origin of these conflicting results.

## 2. Methods

The simulation data on GB migration velocities presented in this paper are all based on Eq. 2 and the calculations were carried out using MATLAB software. In Fig. 1, the curvature is distributed in the range of $0 < \kappa \leq 0.1$ μm$^{-1}$; the parameters for GB effective step energy take values of $\varepsilon^* = 0, 1 \times 10^{-16}, 1 \times 10^{-14}$ and $1 \times 10^{-13}$ J/m, with GB energy fixed at $\gamma = 1$ J/m$^2$; activation energy $Q = 50$ KJ/mol, and atomic step height $l: 2 \times 10^{-10}$ m, with a temperature

T=1073K. Each curvature in Fig.1a-1c has 500 calculated points and these data points are evenly spaced within the corresponding $n$ value range. GB migration velocities for each curvature $\kappa$ in Fig. 1d are taken from the experimental data within the range of $\kappa \sim \kappa + \Delta\kappa$ (where $\Delta\kappa = 0.001$ μm$^{-1}$).

In Fig. 2, the curvature is distributed in the range of $0 < \kappa \leq 0.5$ μm$^{-1}$ which curvature range references the experimental results from literature in Ref. 15. The parameters for GB step energy take values of $\varepsilon^2 = 0$, $1.6 \times 10^{-23}$, $1 \times 10^{-23}$, $2 \times 10^{-24}$, $2 \times 10^{-25}$ and $2 \times 10^{-26}$ (J/m)². The uniform GB energy and random GB energy values is set at $\gamma = 1$ J/m² and 0.18~1.81 J/m², respectively. The range for the random GB energy is limited to a typical order of magnitude. The activation energy $Q = 100$ KJ/mol, atomic step height $l$: $2 \times 10^{-10}$ m, and temperature T=1073K. More than 600,000 GBs were calculated for each $\varepsilon$ with a curvature step of $1 \times 10^{-5}$ μm$^{-1}$. The computed number of GBs was designed to decrease with the increase of curvature, according to the decreasing $n$ value with curvature and the existing experimental observations in the literature. In Fig. 3, the average migration velocities were calculated from the same original data as Fig. 2. The data were categorized into continuous curvature groups.

## 3. Results and Discussions

### 3.1 GB migration model for simulations

In contrast to the extensively studies classical reaction rate theory, GB migration formula in general growth theory have received little attention. While general grain growth theory has been proven to effectively explain general behaviors of grain growth, the properties of GB migration formula within this theory have not been systematically investigated. For instance, the statistical correlation and its evolution law between GB migration velocity and curvature in complex polycrystalline systems remains unexplored. The novel curvature-driven GB migration formula in the general grain growth theory is as follow [14],

$$v = M\gamma\kappa(n-1)e^{\left(-C \cdot \frac{\varepsilon^*}{T(n-1)\kappa}\right)} \qquad (2)$$

where $n$ is a dimensionless variable that equal to the ratio of surface curvature of grains on both sides of a GB, *i.e.* $n = \kappa_a/\kappa$. $C = \pi/2K_B$ and $T$ are the constant coefficient and temperature. $\varepsilon^*$, the effective GB step energy defined as $\varepsilon^* = \varepsilon^2/l\gamma$, has units of the free energy per unit length like $\varepsilon$ and strongly correlates with GB microstructural characteristics and temperature. Here $\varepsilon$ represents the GB step energy reflecting the atomic roughness of the boundary and $l$ is the height of the new layer. Compared with Eq.1 in the classic reaction rate theory, Eq.2 introduces two new variables (namely temperature-independent variable $n$ and temperature-dependent variable $\varepsilon^*$) and an additional exponential term. These two newly introduced variables are, respectively, directly linked to the curvature distribution (or grain size distribution) and GB characteristics within a polycrystalline system. In contrast to the single correspondence between GB migration velocity and curvature described by classic models, Eq.2 reveals that GB with the same curvature in a polycrystalline system may have a series of different migration velocities due to the existence of curvature-distribution-related variable $n$, which prediction has been validated by recent observations from 3D GB migration experiments of polycrystalline Ni [15], Fe [23] and SrTiO3 ceramics [25]. The variable $\varepsilon^*$, analogous to GB energy, is strongly influenced by microstructural characteristics of GBs and temperature; however, it demonstrates markedly higher sensitivity to changes in these factors than GB energy. The introduction of $\varepsilon^*$ not only sensitively reflects the differences between GB structures (such as GB anisotropy) and their changes with temperature (such as GB roughening transitions), but also avoids incorporating complex parameters describing the geometric structure of GBs into the equation. Since $\varepsilon^*$ is an intrinsic property of GBs, Eq.2 is applicable to diverse types of GBs and polycrystalline materials, thereby facilitating the derivation of a grain growth equation that describes general grain growth behavior, which has been corroborated by experimental observations [20]. In this work, we explore the properties of Eq.2 and uncover the underlying cause of the aforementioned conflicting results observed in 3D nondestructive grain-growth studies. Furthermore, the formula of Eq.2 is furtherly generalized to describe GB migration driven by factors besides curvature, such as the stored energy resulting from defects or strain.

GB migration in a polycrystalline system strongly depends on three variables: $\kappa$, $n$ and $\varepsilon^*$ in Eq.2, according to the general theory of grain growth [14]. The variable $\varepsilon^*$ is strongly influenced by temperature and GB characteristics (involving GB structures and compositions) in polycrystalline materials. According to the empirical equation of $\varepsilon$, namely, $\varepsilon \approx \exp\left[\frac{-C}{\sqrt{T_R-T}}\right]$ (where $T$ and $T_R$ are the temperature and the roughening temperature, respectively, and $C$ is a constant) [26], GB roughening transition occurs at $T_R$ corresponding to the condition of $\varepsilon_i^* = 0$ in Eq.2. The increase in temperature can result in the decrease in $\varepsilon^*$ until zero that is accompanied by GB roughening transition. According to the definition, the range of $n$ value is determined by the distribution of curvatures in polycrystalline materials. Since the curvature of grain is correlated with grain size [12], each curvature can statistically correspond to a different range of $n$ values due to grain size distribution in a polycrystalline system. The smaller curvature, statistically corresponding to large grains, has a greater range of $n$ values. The curvature ranges observed in previous 3D GB migration experiments in the literature were primarily between 0 and 0.1 μm-1 due to the limitation of resolution. The numerical analysis is mainly grounded on this curvature range, i.e. $0 < \kappa \leq 0.1\ \mu m^{-1}$. The parameters in this numerical analysis used typical values, see the detailed in the simulation section.

**3.2** The relationship between GB migration velocity and curvature

According to Eq. 2, the relationship between GB migration velocities and curvature changes with $\varepsilon^*$ as shown in Fig.1. As $\varepsilon^*$ decreases with rising temperature or changing GB characteristics, GB migrations shift from complete stagnation to a situation where parts of GBs with smaller curvatures occur detectable migration due to greater $n$ values. Meanwhile, the large-curvature GBs and small-curvature GBs with smaller $n$ values still maintain migration stagnation as shown in Fig. 1A. In this case, small-curvature GBs have migration advantage, aligning with prediction from classical grain growth models. As $\varepsilon^*$ continues to decrease, GBs of all curvatures except the maximum may migrate, while their counterparts with smaller $n$ values remain stagnation. These migration velocities of GBs with the same curvature can span several orders of magnitude and include both migrating and stationary GBs, as showed in Fig. 1B. The occurrence of AGG was attributed to the coexistence of migrating and stationary GBs [20]. Unexpectedly, most of these GBs have similar migration velocity ranges, despite the wide variation in $n$ values for GBs with different curvatures. Only at the GBs very close to the maximum curvature (equivalently the smallest grain size) in the polycrystalline system, the upper limit of their migration velocity range significantly decreases, where the corresponding $n$ values are very small. In this case, smaller-curvature GBs show slow migration velocity increases across a wide range of $n$ values following an initial rapid increase at small $n$ values. As the GB curvature increases, the migration velocities of GBs with the same curvature rapidly increase with rising $n$ values. In GBs with larger curvature, even a slight change in the $n$ value can lead to several times difference in migration velocity, which may explain the experimental observation that very similar GBs in annealed polycrystalline Fe have migration velocities differing by a factor of nine [21]. These features lead to a similar range of migration velocities for GBs with varying curvatures in a polycrystalline system. When $\varepsilon^*$ continues to decrease to zero, the Eq.2 reduces to a form similar to Eq.1 as follow,

$$v = M\gamma(n-1)\kappa \qquad (3)$$

In this context, the migration velocity of a specific GB is linearly related to its curvature, with $M\gamma(n-1)$ as the constant coefficient. For example, in bicrystal GBs with a constant $n$ value (*i.e.*, $n = -1$ due to $\kappa_a = -\kappa$ in bicrystal GB), the GB migration experiments have demonstrated a linear relationship between migration velocity and curvature [11]. For $\varepsilon^* = 0$, the migration velocity distribution with curvature in Fig. 1c is similar to that of case in Fig. 1b. However, the range of GB migration velocities in this case is narrowly distributed within about 3 to 4 orders of magnitude, which is much smaller than the previous cases of $\varepsilon^* \neq 0$. Moreover, for $\varepsilon^* = 0$, there are no stagnant GBs and the grains are in the normal growth stage [14], unlike the case of $\varepsilon^* \neq 0$. The theoretical predictions of the velocity distribution and the similar velocity range of GBs across different curvatures in Fig. 1B and 1C can be well confirmed by the experimental observations (Fig. 1D) in annealed polycrystalline Ni [15], Fe [23] and SrTiO$_3$ [25]. Moreover, the experiment, illustrated in Fig. 1D, also observed both migrating and stagnant GBs at the same curvature and a significant decrease in the number of migrating GBs with increasing curvature, which is attributed to the existence of different $n$ values for each curvature as shown in Fig.1B. In summary, Eq. 2 can provide a comprehensive picture of GB migration in polycrystalline materials. It demonstrates that the curvature-related dimensionless variable n determines both the velocity distribution of GBs with varying curvatures and its evolution over time. Additionally, the range of GB migration velocities is strongly influenced by the temperature-dependent $\varepsilon^*$.

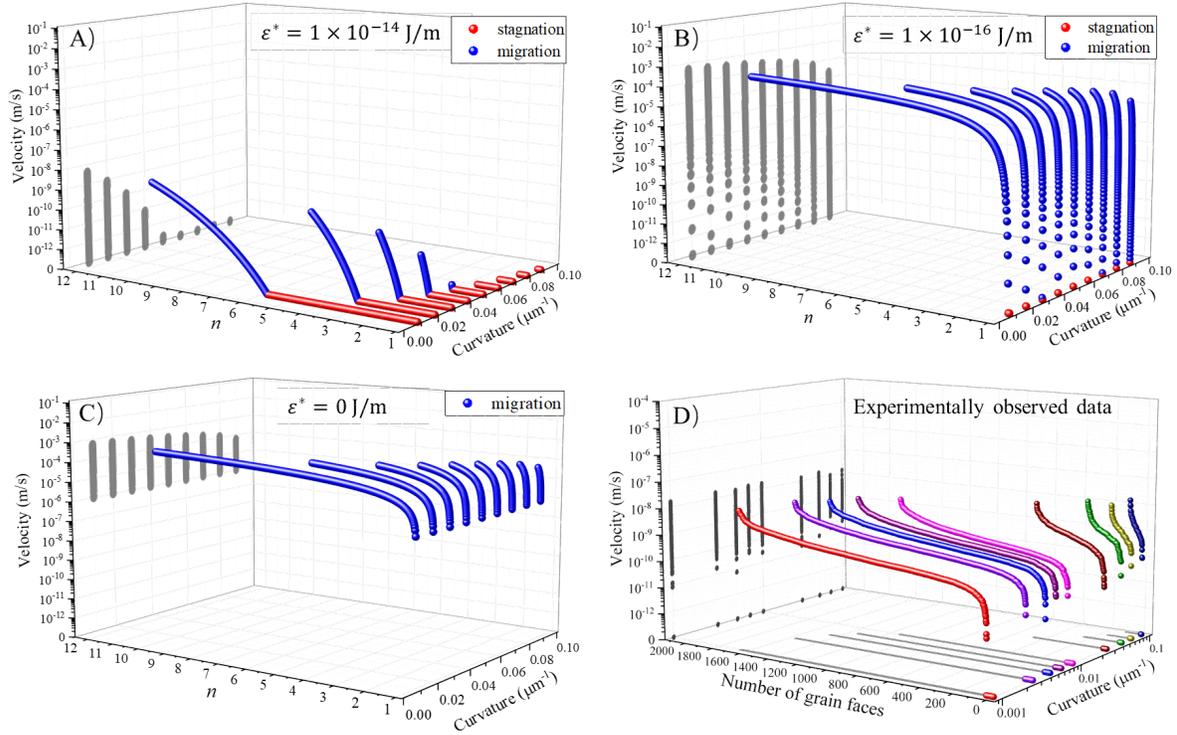

Fig. 1. The evolution of GB velocity distribution with GB curvature and GB efficient step energy ($\varepsilon^*$). **A)-C)** simulation results of different $\varepsilon^*$ values, ($\varepsilon^* \neq 0$ and $\varepsilon^* = 0$ are related to AGG and NGG, respectively), **D)** the experimentally observed data obtained from the Ref.15. **B)-D)** show that the range of GB velocity remain consistent across different curvatures and is strongly influenced by $\varepsilon^*$ in a polycrystalline system (see the gray line projection on the left).

**3.3** Effect of GB energy on GB migration velocity

The effect of GB energy on GB migration velocity and related grain growth behaviors (such as AGG and GGS) has always attracted wide attention and been extensively studied [16,27]. However, the role of GB energy, determined by its five macroscopic degrees of freedom, on GB migration has been controversial. Anisotropic GB energy is often believed to cause AGG. Recent 3D GB migration experiments in polycrystalline Fe [21,23] and Ni [15] yielding differing conclusions about the role of GB energy. In 3D GB migration experiments, Fe sample annealed at 600°C and Ni sample annealed at 800°C showed that GB energy significantly influenced GB migration velocity [15,23]. In contrast, polycrystalline Fe annealed at 800°C exhibited no strong statistical correlation between the five macroscopic degrees of freedom of GBs and reduced mobility, indicating that GB energy has little impact on GB migration velocity [21]. In order to explore this fundamental issue, the distribution of GB migration velocity as a function of GB step energy ($\varepsilon$) and curvature was computed using Eq.2 under two conditions: random GB energy (range $\gamma = 0.18\sim1.81$ J/m) and uniform GB energy (typical value $\gamma =1$ J/m). In the curvature range of $0 < \kappa \leq 0.1$ $\mu m^{-1}$, more than 600,000 GBs were calculated for each $\varepsilon$ with a curvature step of $1 \times 10^{-5}$ $\mu m^{-1}$. The computed number of GB was designed to decrease with the increase of curvature, based on the corresponding $n$ value range and the existing experimental observations.

The computed results in Fig. 2 demonstrate the generally accepted view that GB energy play an important role on GB migration. Unexpectedly, the extent of GB energy's impact on GB migration depends on the value of $\varepsilon$. When $\varepsilon$ is not zero, anisotropic GB energies were revealed to statistically lead to a wider velocity range and a steeper distribution curve of GB migration velocity than that of uniform GB energy in the polycrystalline system as shown in Fig. 2A-2B. These trends become more pronounced as $\varepsilon$ increases. However, these differences caused by anisotropic GB energies are statistically insignificant when $\varepsilon$ is equal to zero, as shown in Fig. 2C. At this point, GB roughening transition occurs, accompanied by a change from AGG to NGG, and Eq. 2 can be simplified into Eq.3 similar in form to the classic Eq. 1. If we consider $M\gamma(n-1)$ in Eq. 3 as the reduced mobility, GB migration velocity is linearly related to its curvature. The value range of $M\gamma(n-1)$ spans 3 to 4 orders of magnitude as shown in Fig. 1c, which theoretical prediction is consistent with the recent 3D GB migration experiment on polycrystalline Fe annealed at 800°C. In this case, the grain growth behavior was found to be approaching or already in the normal growth stage [22], indicating GB migration was very close to or align with the description in Eq.3. Furthermore, the reduced mobility, as the coefficient between GB velocity and curvature, was experimentally found to span about 3 orders of magnitude [21], which aligns with theoretical predictions for the range of $M\gamma(n-1)$ mentioned above. Since the difference in anisotropic GB energy usually doesn't exceed an order of magnitude [28], the $n$ value can be significantly larger. Therefore, when $\varepsilon$ is zero, the variable $n$ has a statistically greater impact on GB migration than anisotropic GB energy. Moreover, as grain growth progresses, the $n$ values change accordingly by definition, which is the root cause of the experimentally observed variation in reduced mobility over time [21].

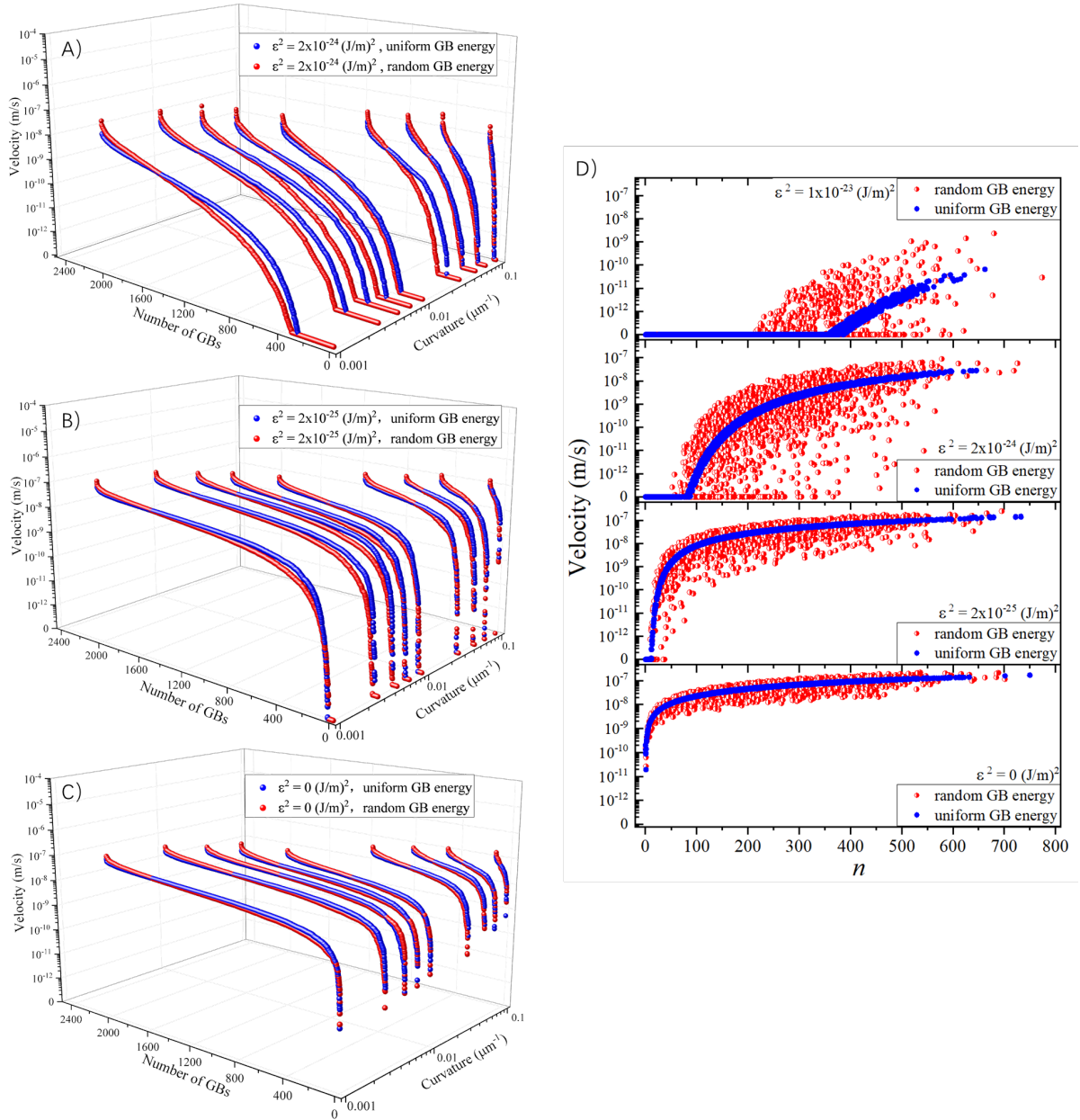

Fig. 2. The influence of GB energy on GB velocity at different GB step energy ($\varepsilon$). **A)**-**C)** the statistical distribution of GB velocities with GB curvatures at different $\varepsilon$ values. **D)** the distribution of GB velocities at a curvature of 0.001 $\mu m^{-1}$ with $n$ values under different $\varepsilon$ values. The results show the impact of GB energy on GB migration velocity significantly decreases as the step energy $\varepsilon$ reduces.

To illustrate the influence of GB energy on the specific GB migration, GB with a curvature of 0.001 $\mu m^{-1}$ is selected for analysis and demonstration as shown in Fig. 2D. Similar to the results in Fig. 2A-2C, the impact of GB energy on specific GB migration velocity also increases with $\varepsilon$, but it is more pronounced than the effects observed in the statistical distribution results. For the case where the $\varepsilon$ is equal to or very close to zero, although the statistical distribution in Fig. 2C indicates minimal influence from GB energy, However, for a specific GB, anisotropic GB energies (typically differing by one order of magnitude) can lead to a tenfold difference in migration velocity. When $\varepsilon$ is non-zero, slight differences in GB energy can even cause variations in GB migration velocity by several orders of magnitude due to the existence of non-linear exponential term in Eq.2, as shown in Fig. 2D. The higher $\varepsilon$, the more pronounced the effect of GB energy anisotropy on the differences in GB migration velocity. Conversely, lower $\varepsilon$ results in smaller variations in migration velocity due to anisotropy. When $\varepsilon$ is zero, the influence of GB energy anisotropy on migration velocity differences is minimized as shown in Fig. 2D. Moreover, the influence of GB energy on GB migration velocity at the same curvature depends on the $n$ value. At initial lower $n$ values, where GB migration velocity increases rapidly with $n$ value, GB energy has a more pronounced effect. Since the temperature-dependent $\varepsilon$ decreases with increasing temperature, the increasing temperature can also reduce the effect of anisotropic GB energies on GB migration velocity. Therefore, the conflicting conclusions regarding GB energy's effect on GB migration observed in 3D GB migration experiments on polycrystalline Fe samples annealed at 600°C [23] and 800°C [21,22], respectively, should be attributed to temperature-induced variations in $\varepsilon$. The Fe sample annealed at 600°C had a higher $\varepsilon$, making the influence of anisotropic GB energy significant. Conversely, the Fe sample annealed at 800°C exhibited a lower or nearly zero $\varepsilon$ as discussed above, resulting in a minimal statistical impact of anisotropic GB energy on migration velocity. In summary, the effect of GB energy on migration is influenced by $\varepsilon$, which varies with temperature and GB characteristics.

**3.4** Distribution of average GB migration velocity

Recent experiments on polycrystalline Fe and Ni revealed that the distribution of the average GB migration velocity shows a trend of widening and diverging with increasing curvature as shown in Fig. 3A, which led to the conclusion that GB migration velocity is not linearly related to GB curvature [15,23]. The reason for this trend has not yet been explained. Based on Eq.2, the average GB migration velocity distributions under different $\varepsilon$ values were numerically simulated and compared with the experimental data from literature as shown in Fig. 3. The simulated results of average GB migration velocity (curvature bin = $0.00001 \mu m^{-1}$) in Fig. 3B is revealed to have a similar diverging trend with the experimental results (also using curvature bin = $0.00001 \mu m^{-1}$) in Fig. 3C. However, when the statistical quantities of GB migration velocity are identical across curvatures as shown in Fig. 3D, the distribution of average migration velocity displays a similar width. This reveals that the experimentally observed trend is not an inherent feature of polycrystalline system, but rather depends on the statistical number of GB migration velocities at specific curvatures. A significantly larger statistical number at the same curvature can result in a narrower average migration velocity distribution. The distribution width of the average GB migration velocity is influenced by the statistical quantity, primarily due to the similar migration velocity range across different curvatures in Fig. 1. Furthermore, as shown in Fig. 3B, the distribution range and values of average GB migration velocity strongly depend on $\varepsilon$. As $\varepsilon$ decreases, the velocity range narrows significantly, and the average velocities may increase by several orders of magnitude.

According to Eq.2, GBs with smaller curvatures have larger $n$ value ranges. Consequently, in actual polycrystalline systems, GBs with smaller curvature tend to migrate more frequently, leading to a higher observed number of migrated GBs. With increased GB curvature, the corresponding $n$ value range decreases, leading to fewer observed migrated GBs in the polycrystalline system. In addition, the tendency of GB energy minimization during microstructural evolution also facilitates the formation of small-curvature GBs in polycrystalline materials. These factors commonly result in the experimentally observed decrease in GB numbers as curvature increases, as shown in Fig.1D and Fig. 3C. The velocities of the fewer high-curvature GBs are randomly scattered along the steeper velocity distribution curve. As a result, with increasing curvature, the average migration velocity of GBs shows greater fluctuations and a divergent trend as illustrated in Fig. 3A-3C. Therefore, the distribution shape of GB migration velocity is rather related to the curvature distribution in the polycrystalline system.

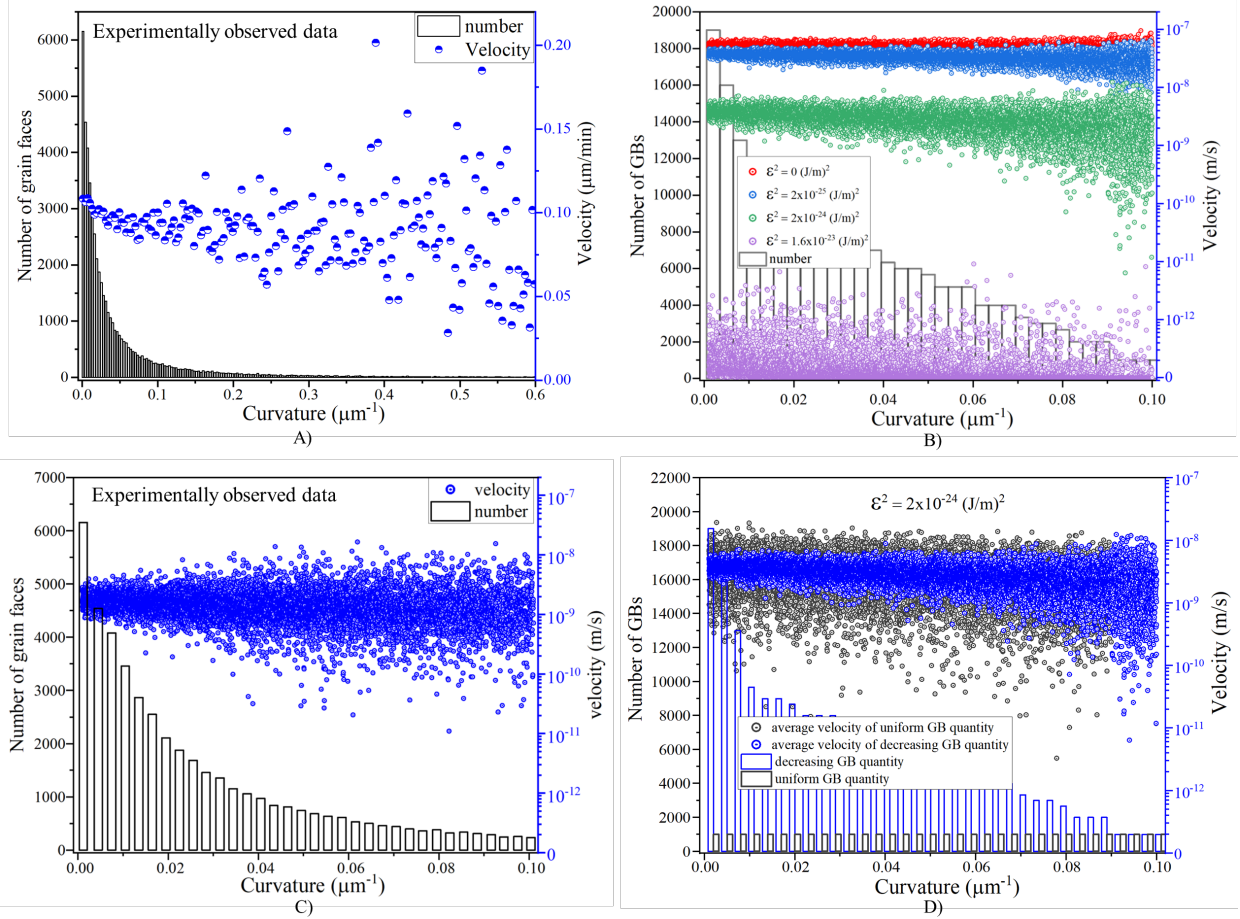

Fig. 3. The distribution of average GB velocity with GB curvature. **A)** the experimentally observed distribution of average GB velocity in polycrystalline nickel, binned by curvature with widths of 0.003 μm$^{-1}$. **B)** the computed distribution of average GB velocity at different $\varepsilon$ values. **C)** the experimentally observed distribution of average GB velocity in polycrystalline nickel, binned by curvature with widths of 0.00001 μm$^{-1}$. **D)** the comparison of the distributions of average GB velocity at a specific $\varepsilon$, computed using two different GB quantity distribution. Experimental data of **B)** and **C)** were obtained from Ref.15. The results show that the trend of the average GB velocity distribution with curvature is correlated with the distribution of the GB quantity in polycrystal.

**3.5** GB migration model for driving force beyond curvature

From the discussion above, it's clear that Eq. 2 effectively explains the current experimental results of 3D GB migration and reveals the source of conflicting conclusions drawn from these experiments. Moreover, it clarifies why experimental results on 3D GB migration deviate from the predictions of the classical GB migration formula (Eq. 1). According to the new GB migration model [14], the process is driven by differences in surface curvature between adjacent grains across GB, making it applicable to various types of GBs, including those containing nanometer-scale amorphous films as well as GBs without such intergranular films. Different structural and chemical compositions at GBs modify the GB step energy and GB energy, thereby influencing GB migration kinetics.

Similarly, in multi-phase systems, grain growth (known as coarsening) is also driven by surface curvature differentials between grains, even when grains are separated by liquid phases or other secondary phases. In this context, the variable $n$ in Eq. 2, representing the ratio of surface curvatures between adjoining grains, can be interpreted as the ratio between the curvature of the growing grain and the mean curvature of its surrounding grains. This implies that the range of $n$ values for growing grains of different sizes in multi-phase systems is significantly narrower than that in single-phase systems. When $\varepsilon^*$ approaches or equals zero (corresponding to the absence of significant abnormal grain growth), the $n$ values among different growing grains in multi-phase systems with mono-modal size distributions exhibit minimal variation, and the distribution of interface migration velocities becomes significantly narrower, as illustrated in Fig. 1C and 3B. Under these conditions, the average interface migration velocity observed in multi-phase systems may demonstrate an approximately linear relationship with curvature. Moreover, as discussed in Section 3.4, deviations from this linear relationship become more pronounced with increasing curvature. These theoretical predictions align well with recent experimental observations of grain growth in Al-Cu alloys [24]. Therefore, Eq. 2 is equally applicable for describing interface migration behavior during grain growth processes in multi-phase systems.

Furthermore, the current formula for GB migration considers only the curvature difference on both sides of the boundary as the driving force. In reality, polycrystalline materials contain various defects (such as dislocations commonly found in metals), which also generate significant stress and strain energy in the lattice. These energies, like the GB curvature difference, can serve as a driving force for the growth of grains, as seen in the recrystallization process of metals. Therefore, Eq. 2 should be expanded into a more general formula for GB migration, so it can describe other stored energies and/or external fields (like magnetic field) as driving forces, including the curvature-driven forces. According to our previous article [14], the general expression of GB migration velocity is as follows:

$$\nu = \lambda w_0 e^{\left(-\frac{\Delta G^*}{K_B T}\right)}[1 - e^{\left(-\frac{\Delta G}{K_B T}\right)}] \tag{4}$$

where $w_0$ and $\Delta G^*$ are the frequency factor and the apparent activation energy (or the energy barrier) of GB migration, respectively. $\lambda$ is the distance of each jump of migration unit. Using the equipartition theorem, the frequency factor can be expressed by the form $w_0 = \frac{K_B T}{h}$. Here, $K_B$, $T$ and $h$ are the Boltzmann's constant, the temperature and the Plank's constant. $\Delta G$ is the driving force for atomic transport, which originates from the difference in the free energy of material on both sides of a GB. The free energy difference can result from surface curvature difference between the adjacent grains and/or from other sources like defects and external fields. Therefore, the difference in the free energy can be written as $\Delta G = \Omega(\Delta P + \Delta G_V)$, where $\Delta P$ is the capillary pressure difference caused by the curvature difference on both sides of the boundary. $\Delta G_V$ is the difference in the free energy per unit volume on both sides of the boundary, which originating from the energy-stored sources (such as defects and elastic deformation) and/or external fields (such as magnetic field). $\Omega$ is the atomic (or atomic cluster) volume. According to the general model [14], the apparent activation energy of GB migration can be written as $\Delta G^* = \Delta G^*_{det} + \Delta G^*_{att}$, where $\Delta G^*_{det}$ is the energy barrier for an atom (or atomic cluster) detaching from the grain surface. The atom-detached process is a thermally activated process, *i.e.* $\Delta G^*_{det} = \Delta Q$ (where $\Delta Q$ is the activation energy of atom detachment). $\Delta G^*_{att}$ is the critical energy barrier for atomic attachment to form a stable layer at GB (Supported materials for more details), given by $\Delta G^*_{att} = \frac{\pi \varepsilon^2}{l \cdot (\Delta P + \Delta G_V)}$. Here, $\varepsilon$ is GB step energy, identical to that in Eq. 2. $l$ and $\pi$ are the height of new layer and constant, respectively. Due to the value of $\Delta G$ usually far smaller than that of $K_B T$, *i.e.* $\Delta G \ll K_B T$, the term of $1 - e^{\left(-\frac{\Delta G}{K_B T}\right)}$ in Eq. 4 is approximately equal to $\frac{\Delta G}{K_B T}$, i.e. $1 - e^{\left(-\frac{\Delta G}{K_B T}\right)} \cong \frac{\Delta G}{K_B T}$. Now, Eq. 4 can be rewritten as:

$$\nu = M(\Delta P + \Delta G_V) \cdot e^{\left(-C \cdot \frac{\varepsilon^2}{T \cdot (\Delta P + \Delta G_V)}\right)} \tag{5}$$

where, $M = \frac{\lambda \Omega}{h} e^{\left(-\frac{\Delta Q}{K_B T}\right)}$ is the temperature-dependent coefficient that can be regarded as GB mobility, analogous to the counterpart in the classical models. $C = \pi/l K_B$ is the constant.

Besides curvature differences, the driving force also includes the stored free energy like defect strain energy in Eq. 5. Therefore, Eq. 5 can describe the growth process of grains during recrystallization of processed metal. When $\Delta G_V \ll \Delta P$, Eq. 5 can be simplified to Eq. 2. In this case, GB migration exhibits the aforementioned characteristics: the distribution trend of GB migration velocities in a polycrystalline system is related to curvature, but the velocity range for GBs with different curvatures is similar and largely independent of GB curvature. The influence of GB energy on migration velocity increases with $\varepsilon$, and so on. When $\Delta G_V \gg \Delta P$, Eq. 5 can be simplified to $v = M \cdot \Delta G_V \cdot e^{\left(-C \cdot \frac{\varepsilon^2}{T \cdot \Delta G_V}\right)}$. In this case, GB migration is independent of GB curvature but still depends on the magnitude of $\varepsilon$ and energy density ($\Delta G_V$). Moreover, when $\Delta G_V$ is sufficiently high and/or $\varepsilon$ is small enough, causing the exponential term in Eq. 5 to approach one, the relationship between $\Delta G_V$ and $v$ becomes approximately linear. This prediction of linear relationship align well with the GB migration experiments in zinc under high magnetic fields (namely high $\Delta G_V$) [29]. Since Eq. 2 exhibits universality across various types of GBs, Eq. 5, as its further extension, is likewise applicable to different GB types and materials. However, its validity still requires further experimental verification.

## 4. Conclusions

The recently-developed GB migration model effectively explains existing experimental results from both bicrystals and polycrystals. GB migration velocity has an exponentially nonlinear relationship with curvature at non-zero GB step energies, but becomes linear at zero GB step energy. The distribution range of GB migration velocities shows minimal correlation with curvature magnitude, but narrows significantly as GB step energy decreases. The distribution converges to 3~4 orders of magnitude as GB step energy approaches zero. The influence of GB energy on GB migration decreases significantly as GB step energy reduces. Finally, this GB migration model can be further expanded to incorporate general stored energy as a driving force, beyond the traditional capillary force.

**CRediT authorship contribution statement**
**Jianfeng Hu:** Conceptualization, Writing – original draft, Writing – review & editing, Visualization, Software, Methodology, Investigation, Formal analysis, Data curation, Funding acquisition.

**Declaration of competing interest**

The authors declare that they have no known competing financial interests or personal relationships that could have appeared to influence the work reported in this paper.

**Acknowledgments**

This work was supported by the National Natural Science Foundation of China (NSFC) under Grant No. 52173223 to J.H. The authors would like to thank Prof. Gregory S. Rohrer for the helpful discussions and his generous sharing of experimental data. The authors would like to thank Xinlei Pan for his assistance in programming.

**Reference**

[1]  Z. Shen, Z. Zhao, H. Peng, and M. Nygren, Formation of tough interlocking microstructures in silicon nitride ceramics by dynamic ripening, Nature **417**, 266 (2002).

[2]  J. Hu, H. Gu, Z. Chen, S. Tan, D. Jiang, and M. Rühle, Core–shell structure from the solution–reprecipitation process in hot-pressed AlN-doped SiC ceramics, Acta Mater. **55**, 5666 (2007).

[3]  X. Zhou, X. Y. Li, and K. Lu, Enhanced thermal stability of nanograined metals below a critical grain size, Science (80-. ). **360**, 526 (2018).

[4]  J. Hu and Z. Shen, Grain growth competition during sintering of SrTiO3 nanocrystals: Ordered coalescence of nanocrystals versus conventional mechanism, Scr. Mater. **194**, 1 (2021).

[5]  Z. Jin, X. Li, and K. Lu, Formation of Stable Schwarz Crystals in Polycrystalline Copper at the Grain Size Limit, Phys. Rev. Lett. **127**, 1 (2021).

[6]  X. Y. Li, Z. H. Jin, X. Zhou, and K. Lu, Constrained minimal-interface structures in polycrystalline copper with extremely fine grains, Science (80-. ). **370**, 831 (2020).

[7]  X. Zhou, X. Li, and K. Lu, Size Dependence of Grain Boundary Migration in Metals under Mechanical Loading, Phys. Rev. Lett. **122**, 126101 (2019).

[8]  L. Jiang, H. Fu, H. Zhang, and J. Xie, Physical mechanism interpretation of polycrystalline metals' yield strength via a data-driven method: A novel Hall–Petch relationship, Acta Mater. **231**, (2022).

[9]  J. Hu and Z. Shen, Grain growth competition during sintering of SrTiO3 nanocrystals: Ordered coalescence of nanocrystals versus conventional mechanism, Scr. Mater. **194**, (2021).

[10] J. Hu and Z. Shen, Grain growth by multiple ordered coalescence of nanocrystals during spark plasma sintering of SrTiO 3 nanopowders, Acta Mater. **60**, 6405 (2012).

[11] G. G. Gottstein and L. S. Shvindlerman, *Grain Boundary Migration in Metals*, 2nd ed. (CRC Press, 2010).

[12] R. D. MacPherson and D. J. Srolovitz, The von Neumann relation generalized to coarsening of three-dimensional microstructures, Nature **446**, 1053 (2007).

[13] M. Hillert, On the theory of normal and abnormal grain growth, Acta Metall. **13**, 227 (1965).

[14] J. Hu, X. Wang, J. Zhang, J. Luo, Z. Zhang, and Z. Shen, A general mechanism of grain growth ─I. Theory, J. Mater. **7**, 1007 (2021).

[15] A. Bhattacharya, Y.-F. Shen, C. M. Hefferan, S. F. Li, J. Lind, R. M. Suter, C. E. Krill, and G. S. Rohrer, Grain boundary velocity and curvature are not correlated in Ni polycrystals, Science (80-. ). **374**, 189 (2021).

[16] G. S. Rohrer, I. Chesser, A. R. Krause, S. K. Naghibzadeh, Z. Xu, K. Dayal, and E. A. Holm, Grain Boundary Migration in Polycrystals, Annu. Rev. Mater. Res. **53**, 347 (2023).

[17] H. Gleiter, Theory of grain boundary migration rate, Acta Mater. **17**, 853 (1969).

[18] L. Zhang, J. Han, Y. Xiang, and D. J. Srolovitz, Equation of Motion for a Grain Boundary, Phys. Rev. Lett. **119**, 1 (2017).

[19] J. Han, S. L. Thomas, and D. J. Srolovitz, Grain-boundary kinetics: A unified approach, Prog. Mater. Sci. **98**, 386 (2018).

[20] J. Hu, J. Zhang, X. Wang, J. Luo, Z. Zhang, and Z. Shen, A general mechanism of grain growth-II: Experimental, J. Mater. **7**, 1014 (2021).

[21] J. Zhang, W. Ludwig, Y. Zhang, H. H. B. Sørensen, D. J. Rowenhorst, A. Yamanaka, P. W. Voorhees, and H. F. Poulsen, Grain boundary mobilities in polycrystals, Acta Mater. **191**, 211 (2020).

[22] J. Zhang, Y. Zhang, W. Ludwig, D. Rowenhorst, P. W. Voorhees, and H. F. Poulsen, Three-dimensional grain growth in pure iron. Part I. statistics on the grain level, Acta Mater. **156**, 76 (2018).

[23] Z. Xu et al., Grain boundary migration in polycrystalline α-Fe, Acta Mater. **264**, (2024).

[24] Z. Xu, J. Sun, J. M. Dake, J. Oddershede, H. Kaur, S. K. Naghibzadeh, C. E. Krill, K. Dayal, and G. S. Rohrer, Grain boundary properties and microstructure evolution in an Al-Cu alloy, Acta Mater. **292**, 121041 (2025).

[25] V. Muralikrishnan et al., Observations of unexpected grain boundary migration in SrTiO3, Scr. Mater. **222**, 115055 (2023).

[26] E. H. Conrad, Surface roughening, melting, and faceting, Prog. Surf. Sci. **39**, 65 (1992).

[27] P. R. Cantwell, T. Frolov, T. J. Rupert, A. R. Krause, C. J. Marvel, G. S. Rohrer, J. M. Rickman, and M. P. Harmer, Grain Boundary Complexion Transitions, Annu. Rev. Mater. Res. **50**, 465 (2020).

[28] S. Ratanaphan, D. L. Olmsted, V. V. Bulatov, E. A. Holm, A. D. Rollett, and G. S. Rohrer, Grain boundary energies in body-centered cubic metals, Acta Mater. **88**, 346 (2015).

[29] D. A. Molodov, C. Günster, and G. Gottstein, Grain boundary motion and grain growth in zinc in a high magnetic field, J. Mater. Sci. **49**, 3875 (2014).